\documentstyle[12pt]{article}
\newcommand{\nn}{\nonumber}
\newcommand{\be}{\begin{equation}}
\newcommand{\ee}{\end{equation}}
\newcommand{\bea}{\begin{eqnarray}}
\newcommand{\eea}{\end{eqnarray}}

\def\bfnabla{\mbox{\boldmath $\nabla$}}

\def\bfsigma{\mbox{\boldmath $\sigma$}}
\def\bfgamma{\mbox{\boldmath $\gamma$}}

\def\bfPi{\mbox{\boldmath $\Pi$}}

\def\al{\alpha}
\def\als{\alpha_{\rm s}}
\def\siml{{\ \lower-1.2pt\vbox{\hbox{\rlap{$<$}\lower6pt\vbox{\hbox{$\sim$}}}}\
 }}

\begin{document}
\begin{titlepage}
\begin{flushright}
\tt
CERN-TH/2003-130 \\
HD-THEP-03-27 \\
IFUM-746-FT
\end{flushright}

\vspace{1cm}
\begin{center}
\begin{Large}
{\bf Poincar\'e invariance constraints on NRQCD and potential NRQCD}
\\[2cm]
\end{Large}
{\large Nora Brambilla\footnote{nora.brambilla@mi.infn.it}}\\
{\it Dipartimento di Fisica, Universit\`a degli Studi di Milano\\
     via Celoria 16, 20133 Milano, Italy \\ and\\}
{\it Institut f\"ur Theoretische Physik, Universit\"at Heidelberg\\
     Philosophenweg 16, 69120 Heidelberg, Germany}
 \\[1cm]
{\large Dieter Gromes\footnote{d.gromes@thphys.uni-heidelberg.de}}\\
{\it Institut f\"ur Theoretische Physik, Universit\"at Heidelberg\\
     Philosophenweg 16, 69120 Heidelberg, Germany}\\[1cm]
{\large Antonio Vairo\footnote{antonio.vairo@cern.ch}}\\
{\it Theory Division CERN, 1211 Geneva 23, Switzerland}\\[1cm]
\end{center}

\vspace{1cm}
\begin{abstract}
\baselineskip=16 pt
{\noindent
We discuss the constraints induced by the algebra of the Poincar\'e
generators on non-relativistic effective field theories.
In the first part we derive some relations among the matching
coefficients of the HQET (and NRQCD), which have been formerly obtained
by use of reparametrization invariance. In the second part we obtain
new constraints on the matching coefficients of pNRQCD.
\vspace{0.5cm}\\ }
\end{abstract}

\end{titlepage}

\vfill\eject

\setcounter{footnote}{0}
\pagenumbering{arabic}

\noindent
\section{Introduction}
For any Poincar\'e invariant theory the generators $H,{\bf P},{\bf J},{\bf K}$
of time translations, space translations, rotations, and Lorentz
transformations satisfy the Poincar\'e algebra:
\begin{eqnarray}
[{\bf P}^i,{\bf P}^j] &=&0, \label{A1}\\
{[{\bf P}^i,H]} &=& 0, \label{A2}\\
{[{\bf J}^i,{\bf P}^j]} &=& i \epsilon_{ijk}{\bf P}^k , \label{A3}\\
{[{\bf J}^i,H]} &=& 0, \label{A4}\\
{[{\bf J}^i,{\bf J}^j]} &=& i \epsilon_{ijk}{\bf J}^k , \label{A5}\\
{[{\bf P}^i,{\bf K}^j]} &=&-i \delta_{ij} H , \label{A6}\\
{[H,{\bf K}^i]} &=&-i {\bf P}^i , \label{A7}\\
{[{\bf J}^i,{\bf K}^j]} &=& i \epsilon_{ijk}{\bf K}^k , \label{A8}\\
{[{\bf K}^i,{\bf K}^j]} &=&-i \epsilon_{ijk}{\bf J}^k . \label{A9}
\end{eqnarray}
It has been pointed out, as early as in Ref. \cite{Dirac}, that the algebra
induces non trivial constraints on the form of the Hamiltonian of
non-relativistic systems where Poincar\'e invariance is no longer explicit.
Indeed, the algebra has been used in the past to constrain the form of
the relativistic corrections to phenomenological potentials \cite{Foldy}.

In this letter by Poincar\'e invariance we mean the explicit realization
of the algebra (\ref{A1})-(\ref{A9}). We study such realization
in some of the modern non-relativistic effective field theories of QCD:
Heavy Quark Effective Theory (HQET) \cite{HQET}, Non Relativistic QCD
(NRQCD) \cite{NRQCD} and potential NRQCD (pNRQCD) \cite{pnrqcd0,pnrqcd}.
One may expect that also in these cases  Poincar\'e invariance induces
non trivial constraints on the form of the interaction. More specifically, one
expects to obtain some exact relations among the matching
coefficients of the considered effective field theory.
This study has never been done before. In the case of the HQET some exact
relations between matching coefficients have been derived using a specific invariance
of the theory, known as reparametrization invariance \cite{maluke,Manohar}.
We will derive in Sec. \ref{secnrqcd} some of these relations, showing
in this way that reparametrization invariance is just a manifestation
of the Poincar\'e invariance of the theory.
In the case where dynamical gauge fields have been integrated out
from pNRQCD, Poincar\'e invariance has been studied in this framework in \cite{bgv}.
There some relations among the potentials have been obtained
that were previously known only from the transformation properties under Lorentz
boost of the representation of the potentials in terms of Wilson loops \cite{poG,poBBP,chen}.
Here we will extend that study to the situation with dynamical gluons.
This is the situation of interest, for instance for weakly coupled heavy
quarkonium states like the $\Upsilon(1S)$.

The letter is organized as follows. In Sec. \ref{secnrqcd}
we illustrate how Poincar\'e invariance (in the sense specified above)
works in NRQCD/HQET. In particular we construct all generators
of the Poincar\'e transformation up to the relevant order  and verify the algebra.
We obtain in this way  some of the constraints already known from reparametrization
invariance. In  Sec. \ref{secpnrqcd} we apply the same machinery to
pNRQCD. After constructing all generators of the Poincar\'e transformations
up to the relevant order, we derive, by imposing the algebra,
some new constraints on the matching coefficients of the theory.
In Sec. \ref{seccon} we summarize our results and discuss some possible
future developments.

\section{NRQCD}
\label{secnrqcd}
After integrating out the hard scale $m$ from QCD, one obtains NRQCD \cite{NRQCD}.
Neglecting operators that involve light-quark fields \cite{ManBauer},
the most general NRQCD Lagrangian density (up to field redefinitions)
for a quark and an antiquark of mass $m$ up to order $1/m^2$
is given by (we display also the term ${\bf D}^4/(8 m^3)$ for further use):
\bea
\nn
&&\hspace{-8mm}
{\cal L}_{\rm NRQCD}=
\\
\nn
&& \hspace{3mm}
\psi^\dagger \Biggl\{ i D_0 -m + c_1 \, {{\bf D}^2\over 2
  m} + c_2 \, {{\bf D}^4\over 8 m^3}
+ c_F\, g {{\bf \bfsigma \cdot B} \over 2 m}
+ c_D^\prime \, g { \left[{\bf D} \cdot, {\bf E} \right] \over 8 m^2}
+ i c_S \, g { {\bf \bfsigma \cdot \left[D \times, E \right] }
\over 8 m^2} \Biggr\} \psi
\\ \nn
&&
+ \chi^\dagger \Biggl\{ i D_0 + m - c_1 \, {{\bf D}^2\over 2 m}
- c_2 \, {{\bf D}^4\over 8 m^3}
- c_F\, g {{\bf \bfsigma \cdot B} \over 2 m}
+ c_D^\prime \, g { \left[{\bf D \cdot, E} \right] \over 8 m^2}
+ i c_S \, g { {\bf \bfsigma \cdot \left[D \times, E\right] }\over 8 m^2}
\Biggr\} \chi
\\ \nn
&&
+ {d_{ss} \over m^2} \psi^{\dag} \psi \chi^{\dag} \chi
+ {d_{sv} \over m^2} \psi^{\dag} {\bfsigma} \psi \cdot \chi^{\dag} {\bfsigma} \chi
+ {d_{vs} \over m^2} \psi^{\dag} {\rm T}^a \psi \chi^{\dag} {\rm T}^a \chi
+ {d_{vv} \over m^2} \psi^{\dag} {\rm T}^a {\bfsigma} \psi \cdot
\chi^{\dag} {\rm T}^a {\bfsigma} \chi
\\
&&
- {1\over 4} F^a_{\mu \nu} F^{a\,\mu \nu}
+ {d_3^\prime\over m^2} g f_{abc}F^a_{\mu\nu} F^b_{\mu\al} F^c_{\nu\al},
\label{lNRQCD}
\eea
where $\psi$ is the Pauli spinor field that annihilates the fermion and $\chi$
is the Pauli spinor field that creates the antifermion,
$i D^0=i\partial^0 -gA^0$, $i{\bf D}=i\bfnabla+g{\bf A}$,
$[{\bf D \cdot, E}]={\bf D \cdot E} - {\bf E \cdot D}$, and
$[{\bf D \times, E}]={\bf D \times E -E \times D}$, ${\bf E}^i = F^{i0}$
and ${\bf B}^i = -\epsilon_{ijk}F^{jk}/2$, $\epsilon_{ijk}$ being
the usual three-dimensional antisymmetric tensor
($({\bf a} \times {\bf b})^i = \epsilon_{ijk} {\bf a}^j {\bf b}^k$).
We display the mass terms, that are usually removed by a field redefinition,
for the further use of the canonical formalism to derive the time-translation generator.
The one-loop expressions in the $\overline{{\rm MS}}$ scheme
for the coefficients $c_F$, $c_D^\prime$, $c_S$ and $d_3^\prime$
can be found in \cite{Manohar} (according to the definitions of $ c_D^\prime$
and $d_3^\prime$ given in \cite{m2}) and for $d_{ij}$ ($i,j=s,v$) in \cite{Match}.

It has been proved in \cite{Manohar} that in the bilinear sector the NRQCD
Lagrangian is equivalent to the Lagrangian of the HQET. Therefore, at the
order at which only the bilinear sector matters (and this will be the case for
the rest of this section), the results that we obtain are valid both  for NRQCD and HQET.
The main reason for introducing here the NRQCD Lagrangian
up to order $1/m^2$, is that it is needed to obtain the pNRQCD
Lagrangian at ${\cal O}(1/m^2)$, which is the subject of Sec. \ref{secpnrqcd}.

\subsection{Canonical quantization}
\label{quantNRQCD}
To be definite we quantize NRQCD in the $A^0 = 0$ gauge \cite{quantization}.
The pairs of canonical variables are $(\psi, i \psi^\dagger)$,
$(\chi, i \chi^\dagger)$ and $({\bf A}_{i\, a},  \bfPi^i_a =
\partial \, {\cal L}_{\rm NRQCD}/$ $\partial \, (\partial_0 {\bf A}_i^a))$.
The physical states $\vert {\rm phys} \rangle$ are constrained
by the Gauss law:
\be
({\bf D}\cdot {\bfPi})^a \vert {\rm phys} \rangle =
g (\psi^\dagger {\rm T}^a \psi + \chi^\dagger {\rm T}^a \chi) \vert {\rm phys} \rangle.
\label{gaussnrqcd}
\ee
The canonical variables satisfy the usual equal time commutation relations:
\bea
[\bfPi^i_a({\bf x},t),{\bf A}^j_b({\bf y},t)] &=&
i \delta_{ij}\delta_{ab} \delta^{(3)}({\bf x}-{\bf y}), \label{comNRQCD1}\\
{[{\bf A}^i_a({\bf x},t),{\bf A}^j_b({\bf y},t)]} &=&
[\bfPi^i_a({\bf x},t),\bfPi^j_b({\bf y},t)] = 0, \label{comNRQCD2}\\
\{\psi_\alpha({\bf x},t),\psi^\dagger_\beta({\bf y},t)\} &=&
\{\chi_\alpha({\bf x},t),\chi^\dagger_\beta({\bf y},t)\} =
\delta_{\alpha\beta} \delta^{(3)}({\bf x}-{\bf y}),  \label{comNRQCD3}\\
\{\psi_\alpha({\bf x},t),\psi_\beta({\bf y},t)\} &=&
\{\psi^\dagger_\alpha({\bf x},t),\psi^\dagger_\beta({\bf y},t)\}  = 0,
\label{comNRQCD4}\\
\{\chi_\alpha({\bf x},t),\chi_\beta({\bf y},t)\} &=&
\{\chi^\dagger_\alpha({\bf x},t),\chi^\dagger_\beta({\bf y},t)\}  = 0.
\label{comNRQCD5}
\eea
All other commutators are zero.
In the following, in order to fulfil the Gauss-law constraint
(\ref{gaussnrqcd}), we will assume the commutators to act
on a space spanned by the physical states.

\subsection{Poincar\'e algebra generators in NRQCD}
The construction of the generators proceeds in the following way.
The generators $H$, ${\bf P}$ and ${\bf J}$ can be derived from the
symmetric energy-momentum tensor (see for instance \cite{weinberg}).
Since translational and rotational invariance
remain  exact symmetries when going to the effective theory,
the transformation properties of the new particle fields
under these symmetries are the same as in the original theory.
The derivation of the Lorentz-boost generators is more problematic,
since the non-relativistic expansion has destroyed the manifest covariance
under boosts. A consistent way to construct ${\bf K}$ is to write down the
most general expression, with some obvious restrictions already included,
containing all operators consistent with its symmetries and to match it to the QCD
Lorentz-boost generator, which is $\displaystyle {\bf K} = -t\, {\bf P}
+ \int d^3x \, {1\over 2} \, \left\{ {\bf x},
{{\bfPi}^a \cdot {\bfPi}^a  + {\bf B}^a \cdot {\bf B}^a \over 2}
+ \bar{\psi}(-i {\bf D}\cdot \bfgamma + m) \psi \right\}$.
This is very much the same procedure that is used in the construction
of the NRQCD Lagrangian. Accordingly, new matching coefficients,
typical of ${\bf K}$, will appear.

Specifically for the Lagrangian (\ref{lNRQCD}) we obtain (${\bf P}$ and ${\bf
  J}$ are exact, $H$
is displayed up to order $1/m^2$, including the kinetic energy up to order
$1/m^3$, and ${\bf K}$ is given up to order $1/m$):
\bea
&& \hspace{-8mm} h \equiv \psi^\dagger \Biggl( m  - c_1 \, {{\bf D}^2\over 2 m}
- c_2 \, {{\bf D}^4\over 8 m^3}
- c_F\, g {{\bf \bfsigma \cdot B} \over 2 m}
- c_D^\prime \, g { \left[{\bf D} \cdot, {\bfPi} \right] \over 8 m^2}
- i c_S \, g { {\bf \bfsigma \cdot \left[D \times, \bfPi \right] }
\over 8 m^2} \Biggr) \psi
\nn \\
&&
+ \chi^\dagger \Biggl( -m  + c_1 \, {{\bf D}^2\over 2 m}
+ c_2 \, {{\bf D}^4\over 8 m^3}
+ c_F\, g {{\bf \bfsigma \cdot B} \over 2 m}
- c_D^\prime \, g { \left[{\bf D \cdot, \bfPi} \right] \over 8 m^2}
- i c_S \, g { {\bf \bfsigma \cdot \left[D \times, \bfPi\right] }\over 8 m^2}
\Biggr) \chi
\nn \\
&&
- {d_{ss} \over m^2} \psi^{\dag} \psi \chi^{\dag} \chi
- {d_{sv} \over m^2} \psi^{\dag} {\bfsigma} \psi \cdot \chi^{\dag} {\bfsigma} \chi
- {d_{vs} \over m^2} \psi^{\dag} {\rm T}^a \psi \chi^{\dag} {\rm T}^a \chi
- {d_{vv} \over m^2} \psi^{\dag} {\rm T}^a {\bfsigma} \psi \cdot
\chi^{\dag} {\rm T}^a {\bfsigma} \chi
\nn \\
&&
+ {{\bfPi}^a \cdot {\bfPi}^a  + {\bf B}^a \cdot {\bf B}^a \over 2}
- {d_3^\prime\over m^2}\, f_{abc} \, g F^a_{\mu\nu} F^b_{\mu\al} F^c_{\nu\al}
\Bigg|_{\hbox{${\bf E}=\bfPi$}} ,
\nn \\
\nn \\
&& \hspace{-8mm} H = \int d^3x \, h ,
\label{HNRQCD} \\
&& \hspace{-8mm} {\bf P} =
\int d^3x \, \left( \psi^\dagger \, (-i{\bf D})  \, \psi
+ \chi^\dagger \, (-i{\bf D}) \, \chi
+ {1\over 2}
\, [{\bfPi}^a \times, {\bf B}^a]\right),
\label{PNRQCD} \\
&& \hspace{-8mm} {\bf J} =
\int d^3x \, \left(\psi^\dagger \left( {\bf x} \times (-i {\bf D})
+ {\bfsigma \over 2}\right)\psi
+ \chi^\dagger \left(  {\bf x} \times (-i {\bf D})  + {\bfsigma \over 2}\right)\chi
\right. \nn\\
&& \hspace{12mm}  \left.
+ {1\over 2}
\, {\bf x} \times [{\bfPi}^a \times, {\bf B}^a]\right),
\label{JNRQCD} \\
&&\hspace{-8mm} {\bf K} =
- t \, {\bf P} + \int d^3x \, {\{ {\bf x}, h \} \over 2}
\nn\\
&&
- k^{(1)} \,
\int d^3x \, \left( {1\over 2m} \psi^\dagger \, {\bfsigma \over 2} \times
(-i{\bf D})  \, \psi
- {1\over 2m} \chi^\dagger \, {\bfsigma \over 2} \times (-i{\bf D})
\, \chi \right),
\label{KNRQCD}
\eea
where $k^{(1)}$ is a matching coefficient specific of ${\bf K}$.

\subsection{Poincar\'e algebra constraints in NRQCD}
\label{secnrqcdcons}
Let us now consider the constraints induced by the Poincar\'e algebra
(\ref{A1})-(\ref{A9}) on the NRQCD generators $H$ and ${\bf K}$.
The constraint $[{\bf P}^i,{\bf K}^j] = -i \delta_{ij} H$
has been already used in Eq. (\ref{KNRQCD}).
Indeed, this commutation relation forces ${\bf K}$ to have the form
$\displaystyle \int d^3x$ $\{ {\bf x}, h({\bf x},t) \}/2$ $+$
translational-invariant terms that depend on ${\bf x}$ only
through the canonical variables.
From $[{\bf K}^i,{\bf K}^j] = -i \epsilon_{ijk}{\bf J}^k$ at ${\cal O}(1/m^0)$
it follows that
\be
-i \epsilon_{ijk} \, (1- k^{(1)}) \, \int d^3x \,
\left( \psi^\dagger {\bfsigma^k \over 2} \psi
+ \chi^\dagger {\bfsigma^k \over 2} \chi\right) =0 \>\> \Rightarrow \>\>
k^{(1)}=1.
\label{constrNRQCD1}
\ee
From $[H,{\bf K}^i] = -i {\bf P}^i$ at  ${\cal O}(1/m^0)$ it follows that
\be
-i \, (1 - c_1) \, \int d^3x \, \left(\psi^\dagger \, (-i{\bf D}^i) \, \psi
+ \chi^\dagger \, (-i{\bf D}^i) \, \chi \right)=0 \>\> \Rightarrow \>\>
c_1=1,
\label{constrNRQCD2}
\ee
and at  ${\cal O}(1/m)$
\be
-i \, (2c_F-c_S-1) \,
\int d^3x \,  \left(\psi^\dagger {\left(\bfsigma \times  g{\bfPi} \right)^i\over 4m} \psi
- \chi^\dagger {\left(\bfsigma \times  g{\bfPi} \right)^i\over 4m} \chi
\right) = 0 \>\> \Rightarrow \>\>
2c_F -c_S -1 =0.
\label{constrNRQCD3}
\ee
Finally from $[H,{\bf K}^i] = -i {\bf P}^i$ at  ${\cal O}(\bfnabla^2 \,
\bfnabla^i /m^2)$ we obtain
\be
(1 - c_2) \, \int d^3x \, \left( \psi^\dagger \, { \bfnabla^2\bfnabla^i \over 2 m^2}  \, \psi
+ \chi^\dagger \, { \bfnabla^2\bfnabla^i \over 2 m^2} \, \chi \right) =0 \>\> \Rightarrow \>\>
c_2=1.
\label{constrNRQCD4}
\ee
All other commutation relations are satisfied at the order we are working.
The constraints (\ref{constrNRQCD2}), (\ref{constrNRQCD3}) and (\ref{constrNRQCD4})
were first derived in the framework of reparametrization invariance
in \cite{maluke,Manohar}.

\section{pNRQCD}
\label{secpnrqcd}
The pNRQCD Lagrangian for a heavy quark-antiquark system is obtained from
NRQCD by integrating out the soft degrees of freedom associated with the scale
of the relative momentum of the two heavy quarks
in the bound state \cite{pnrqcd0,pnrqcd}. The name pNRQCD has been used
in the literature to identify effective field theories with different degrees
of freedom. Here we call pNRQCD the effective field theory that can be
obtained from NRQCD by perturbative matching and contains, as degrees of
freedom, the quark-antiquark field (that can be split into
a colour singlet ${\rm S} = { S 1\!\!{\rm l}_c /
\sqrt{N_c}}$ and a colour  octet ${\rm O} = O^a { {\rm T}^a / \sqrt{T_F}}$ component)
and (ultrasoft) gluons. The fields $S$ and $O^a$ are functions of
$({\bf X},t)$ and  ${\bf x}$, where ${\bf X} = ({\bf x}_1+{\bf x}_2)/2$
is the centre-of-mass coordinate and ${\bf x}={\bf x}_1-{\bf x}_2$ the
relative coordinate, with ${\bf x}_1$ (${\bf x}_2$) the
coordinates of the quark (antiquark). The coordinate ${\bf x}$ plays the role
of a continuous parameter, which specifies different fields.
All the gauge fields have been multipole expanded around the centre-of-mass
and depend on $({\bf X},t)$: $F^{\mu \nu \, a} = F^{\mu \nu \, a}({\bf X},t)$
and $iD_\mu {\rm O} = i \partial_\mu {\rm O} - g [A_\mu({\bf X},t),{\rm O}]$.
The terms in the pNRQCD Lagrangian are organized by powers
in the $1/m$ and $x$ expansions. We will indicate a generic term
of order $x^j/m^i$ as $h_{\varphi\phi}^{(i,j)}$,
where $\varphi,\phi \in \{S, O \}$.
Notice that the derivative $\bfnabla_x$ counts like $x^{-1}$
and, therefore, contributes with a negative power to the second index
of $h_{\varphi\phi}^{(i,j)}$.

We aim at verifying the Poincar\'e algebra at order $x/m^0$ and $x^0/m$.
For this purpose we need the pNRQCD Lagrangian at order $x^2/m^0$
(obtained in \cite{pnrqcd0,decay}), as well as at order $x^0/m$,
$(x/m) \, {\bf P}_X$ and $(x^0/m^2) \, {\bf P}_X$ (derived in this work).
We include all local operators of the correct transformation properties
and the appropriate dimensions, moreover we assume that, apart from the
kinetic energy, momentum and spin operators are suppressed by a factor $1/m$
each. To the above order the pNRQCD Lagrangian density is given by
(as in NRQCD, we display the mass terms, that could be removed by a
field redefinition, for the further use of the canonical formalism
to derive the time-translation generator):
\bea
&& \hspace{-8mm} {\cal L}_{\rm pNRQCD} = \int d^3x \, {\rm Tr} \,\Biggl\{
  {\rm S}^\dagger \left( i\partial_0 - 2m - h_S \right) {\rm S}
+ {\rm O}^\dagger \left( iD_0 -2m - h_O \right) {\rm O}
\nn\\
&&\qquad
- \left[ ({\rm S}^\dagger h_{SO} {\rm O} + {\rm H.C.}) + {\rm C.C.} \right] \;
- \left[ {\rm O}^\dagger h_{OO} {\rm O} + {\rm C.C.}  \right]  \;
- \left[ {\rm O}^\dagger h_{OO}^A {\rm O}  h_{OO}^B + {\rm C.C.}  \right]
\Biggr\}
\nn \\
&&\qquad
- {1\over 4} F^a_{\mu \nu} F^{a\,\mu \nu},
\\
\nn
\\
&& \hspace{-8mm} h_\varphi =
\{c_{\varphi}^{(1,-2)}(x), {{\bf p}_x^2 \over 2m} \}
+ c_{\varphi}^{(1,0)}(x) {{\bf P}_X^2\over 4m}
+ V_\varphi^{(0)}(x) + {V^{(1)}_\varphi(x)\over m} + {V^{(2)}_\varphi\over
  m^2}, \qquad {\rm with}
\\
&& V_\varphi^{(2)} =
V_{r\,\varphi}(x)
+ {1\over 8}\{ {\bf P}_X^2, V_{{\bf p}^2\,\varphi a}(x)\}
+ {1\over 2}\{ {\bf p}_x^2, V_{{\bf p}^2\,\varphi b}(x)\}
\nn\\
&&\qquad
+ {({\bf x}\times {\bf P}_X)^2\over 4 x^2} V_{{\bf L}^2\,\varphi a}(x)
+ {({\bf x}\times {\bf p}_x)^2\over x^2} V_{{\bf L}^2\,\varphi b}(x)
\nn\\
&&\qquad
+ {({\bf x}\times {\bf P}_X)\cdot (\bfsigma^{(1)} - \bfsigma^{(2)}) \over 4}
V_{LS\,\varphi a}(x)
+ {({\bf x}\times {\bf p}_x)\cdot (\bfsigma^{(1)} + \bfsigma^{(2)}) \over 2}
V_{LS\,\varphi b}(x)
\nn\\
&&\qquad
+ {1\over 4}V_{S^2\,\varphi}(x)\, \bfsigma^{(1)}\cdot \bfsigma^{(2)}
+ {V_{{\bf S}_{12}\,\varphi}(x) \over x^2} \, \left(3 \, {{\bf x}}\cdot {\bfsigma}^{(1)}
\,{{\bf x}}\cdot {\bfsigma}^{(2)}  - {\bfsigma}^{(1)}\cdot{\bfsigma}^{(2)}\right)
,
\\
\nn
\\
&& \hspace{-8mm} h_{\varphi\phi} =
h_{\varphi\phi}^{(0,1)} + h_{\varphi\phi}^{(0,2)}
+ h_{\varphi\phi}^{(1,0)} + h_{\varphi\phi}^{(1,1)}({\bf P}_X)
+ h_{\varphi\phi}^{(2,0)}({\bf P}_X),
\qquad {\rm with}
\\
&&  h_{\varphi\phi}^{(0,1)} =
- {V_{\varphi\phi}^{(0,1)}(x) \over 2} {\bf x}\cdot g {\bf E},
\\
&&  h_{\varphi\phi}^{(0,2)} =
- {V_{\varphi\phi a}^{(0,2)}(x) \over 8} {\bf x}^i{\bf x}^j ({\bf D}^i g {\bf
  E}^j)
- {V_{\varphi\phi b}^{(0,2)}(x) \over 8} {\bf x}^2 ({\bf D} \cdot  g {\bf E}),
\\
&&  h_{\varphi\phi}^{(1,0)} =
{1\over 8m} \, V_{\varphi\phi a}^{(1,0)}(x)\,\{ {\bf p}_x \cdot ,{\bf x}\times g {\bf B} \}
-{c_F \over 2m} \, V_{\varphi\phi b}^{(1,0)}(x) \, \bfsigma^{(1)}\cdot g {\bf B}
\nn\\
&&\qquad
-{1\over 2m} \, {V_{\varphi\phi c}^{(1,0)}(x) \over x^2}
\, ({\bf x}\cdot\bfsigma^{(1)})\, ({\bf x}\cdot g {\bf B})
- {1\over m}{V_{\varphi\phi d}^{(1,0)}(x) \over 2 x} {\bf x}\cdot g {\bf E},
\\
&&  h_{\varphi\phi}^{(1,1)}({\bf P}_X) =
{1\over 8m} \, V_{\varphi\phi}^{(1,1)}(x) \,
\{ {\bf P}_X \cdot ,{\bf x}\times g {\bf B} \},
\\
&&  h_{\varphi\phi}^{(2,0)}({\bf P}_X) =
{c_s \over 16m^2} V_{\varphi\phi a}^{(2,0)}(x) \,
\bfsigma^{(1)}\cdot [{\bf P}_X \times, g {\bf E}]
\nn\\
&&\qquad
+ {1\over 16m^2} {V_{\varphi\phi b'}^{(2,0)}(x) \over x^2} \,
({\bf x}\cdot\bfsigma^{(1)}) \,
\{ {\bf P}_X \cdot, (g {\bf E} \times {\bf x}) \}
\nn\\
&&\qquad
+ {1\over 16m^2} {V_{\varphi\phi b''}^{(2,0)}(x) \over x^2} \,
\{ ({\bf x}\cdot g {\bf E}),
{\bf P}_X \cdot ({\bf x} \times \bfsigma^{(1)})\}
\nn\\
&&\qquad
+ {1\over 16m^2} {V_{\varphi\phi b'''}^{(2,0)}(x) \over x^2} \,
\{({\bf x}\cdot{\bf P}_X),
\bfsigma^{(1)} \cdot ({\bf x} \times g {\bf E}) \}
\nn\\
&&\qquad
+ {1\over 16m^2} \,
\{({\bf p}_x\cdot{\bf P}_X), V_{\varphi\phi c'}^{(2,0)}(x)
({\bf x} \cdot g {\bf E}) \}
\nn\\
&&\qquad
+ {1\over 16m^2} \,
\{{\bf p}_x^i {\bf P}_X^j, V_{\varphi\phi c''}^{(2,0)}(x)
\,{\bf x}^j \,g {\bf E}^i \}
\nn\\
&&\qquad
+ {1\over 16m^2} \,
\{{\bf p}_x^i {\bf P}_X^j, V_{\varphi\phi c'''}^{(2,0)}(x)
\, {\bf x}^i \, g {\bf E}^j \}
\nn\\
&&\qquad
+ {1\over 16m^2} \,
\{{\bf p}_x^i {\bf P}_X^j, {V_{\varphi\phi d}^{(2,0)}(x)\over x^2}  \,
{\bf x}^i \, {\bf x}^j \, ({\bf x} \cdot g {\bf E}) \}
\nn\\
&&\qquad
+ {1\over 8m^2} \, {V_{\varphi\phi e}^{(2,0)}(x) \over x}\,
\{ {\bf P}_X \cdot ,{\bf x}\times g {\bf B} \},
\\
\nn
\\
&& \hspace{-8mm} {\rm O}^\dagger h_{OO}^A {\rm O} h_{OO}^B  =
{\rm O}^\dagger h_{OO}^{A\,(1,0)} {\rm O} h_{OO}^{B\,(1,0)}
+ {\rm O}^\dagger h_{OO}^{A\,(2,0)}{\rm O} h_{OO}^{B\,(2,0)}({\bf P}_X),
\qquad {\rm with}
\\
&& {\rm O}^\dagger h_{OO}^{A\,(1,0)} {\rm O} h_{OO}^{B\,(1,0)} =
- {1\over 2m} V_{O\otimes Ob}^{(1,0)}(x) \, {\rm O}^\dagger \bfsigma^{(1)}
\cdot {\rm O}  g{\bf B}
\nn\\
&&\qquad
- {1\over 2m} {V_{O\otimes Oc}^{(1,0)}(x) \over x^2}
\, {\rm O}^\dagger ({\bf x} \cdot \bfsigma^{(1)})
{\rm O}   ({\bf x}\cdot g{\bf B}),
\\
&& {\rm O}^\dagger h_{OO}^{A\,(2,0)} {\rm O} h_{OO}^{B\,(2,0)}({\bf P}_X) =
{1\over 16m^2} V_{O\otimes O a}^{(2,0)}(x) \,[ {\rm O}^\dagger
\bfsigma^{(1)}\cdot ({\bf P}_X {\rm O} \times,  g {\bf E})]
\nn\\
&&\qquad
+ {1\over 16m^2} {V_{O\otimes O b'}^{(2,0)}(x) \over x^2} \, \{ {\rm O}^\dagger
({\bf x}\cdot\bfsigma^{(1)}) \,
 {\bf P}_X  {\rm O}\cdot, (g {\bf E} \times {\bf x}) \}
\nn\\
&&\qquad
+ {1\over 16m^2} {V_{O\otimes O b''}^{(2,0)}(x) \over x^2} \,\{ {\rm O}^\dagger
{\bf P}_X \cdot ({\bf x} \times \bfsigma^{(1)}) {\rm O},
({\bf x}\cdot g {\bf E})\}
\nn\\
&&\qquad
+ {1\over 16m^2} {V_{O\otimes O b'''}^{(2,0)}(x) \over x^2} \,\{ {\rm O}^\dagger
({\bf x}\cdot{\bf P}_X) \bfsigma^{(1)} {\rm O}
\cdot,  ({\bf x} \times g {\bf E}) \}.
\eea
We have defined ${\bf P}_X = -i{\bf D}_X$ and ${\bf p}_x = -i\bfnabla_x
= ({\bf p}_1 - {\bf p}_2)/2$;
when acting on a singlet field ${\bf P}_X$ reduces to $-i\bfnabla_X
= {\bf p}_1 + {\bf p}_2$, ${\bf p}_j$ being $-i \bfnabla_{x_j}$.
The abbreviation C.C. stands for charge conjugation, H.C. for Hermitian conjugation.
For the potential $V^{(2)}$ we have used a notation close to the
traditional one of \cite{m2}: we indicate momentum- and spin-independent
potentials with $V_r$, momentum-dependent potentials with
$V_{{\bf p}^2}$ and $V_{{\bf L}^2}$, spin-orbit potentials with
$V_{LS}$, spin-spin potentials with $V_{S^2}$ and spin-tensor potentials
with $V_{{\bf S}_{12}}$. The letters $a$, $b$, ... that appear
in some of the matching coefficients are used
to label the different kinds of operators.
For $h^{(1,1)}$ and  $h^{(2,0)}$ only the ${\bf P}_X$-dependent terms are displayed.

Under color trace all displayed terms of the type $h_{SS}^{(i,j)}$ vanish.
Due to charge conjugation the terms proportional to $V_{SOa}^{(0,2)}$,
$V_{SOb}^{(0,2)}$, $V_{SOa}^{(1,0)}$, $V_{SOc'}^{(2,0)}$, $V_{SOc''}^{(2,0)}$,
$V_{SOc'''}^{(2,0)}$ and $V_{SOd}^{(2,0)}$ vanish.
The matching coefficients $c_{\varphi}^{(1,-2)}$, $c_{\varphi}^{(1,0)}$,
$V_{\varphi\phi}^{(0,1)}$, $V_{OO a}^{(0,2)}$,
$V_{OOa}^{(1,0)}$,  $V_{\varphi\phi b}^{(1,0)}$, $V_{\varphi\phi}^{(1,1)}$
and $V_{\varphi\phi a}^{(2,0)}$ are equal to $1$ at tree level.
All other matching coefficients are zero at ${\cal O} (\als^0)$.
The tree-level matching has been performed by multipole expanding the NRQCD
Lagrangian (\ref{lNRQCD}) and projecting on singlet and octet two-particles states.

\subsection{Canonical quantization}
\label{quantpNRQCD}
The canonical variables and their conjugates are
$(S, i S^\dagger)$, $(O_a, i O^\dagger_a)$, and $({\bf A}_{i\,a},\bfPi^i_a =
\partial \, {\cal L}_{\rm pNRQCD} /\partial \, (\partial_0 {\bf A}_i^a)$).
The physical states $\vert {\rm phys} \rangle$ are constrained to satisfy
the Gauss law:
\be
({\bf D}\cdot{\bfPi})^a\vert {\rm phys} \rangle = \int d^3x \,
{\rm Tr} \left\{  {\rm O}^\dagger  [g {\rm T}^a, \,{\rm O}] \right\}\vert {\rm phys} \rangle.
\label{gauss}
\ee
The canonical commutation relations are
\bea
[\bfPi^i_a({\bf X},t),{\bf A}^j_b({\bf Y},t)] &=&
i \delta_{ab} \delta_{ij}\delta^{(3)}({\bf X}-{\bf Y}),
\\
{[}S({\bf x},{\bf X},t),S^\dagger({\bf y},{\bf Y},t)] &=&
\delta^{(3)}({\bf x}-{\bf y})\delta^{(3)}({\bf X}-{\bf Y}),
\\
{[}O_a({\bf x},{\bf X},t),O^\dagger_b({\bf y},{\bf Y},t)] &=&
 \delta_{ab}\delta^{(3)}({\bf x}-{\bf y})\delta^{(3)}({\bf X}-{\bf Y}),
\eea
all the other commutators are zero.
As in the case of NRQCD, in order to fulfil the Gauss-law constraint
(\ref{gauss}), we will assume the Poincar\'e algebra commutators
to act on a space spanned by the physical states.

\subsection{Poincar\'e algebra generators in pNRQCD}
As in the case of NRQCD, since translational and rotational invariance are
exact symmetries of the effective theory, the generators $H$, ${\bf P}$
and ${\bf J}$ can be derived from the symmetric energy-momentum tensor.
The pNRQCD Lorentz-boost generators ${\bf K}$ can be derived by
writing down the most general expression, with some obvious restrictions already included,
containing all operators consistent with its symmetries and
by matching it to the NRQCD Lorentz-boost generator (\ref{KNRQCD}).

In our case we obtain at order $x^2/m^0$, $x^0/m$,
$(x/m) \, {\bf P}_X$ and  $(x^0/m^2) \, {\bf P}_X$
(and dropping terms involving 4 matter fields that show up at ${\cal
  O}(x^2/m^0)$, but shall not affect our results):
\bea
&& \hspace{-8mm}
h \equiv  {{\bfPi}^{a\,2} + {\bf B}^{a\,2} \over 2}
+ \int d^3x \, {\rm Tr} \,\Biggl\{
  {\rm S}^\dagger (2m + h_S) {\rm S}
+ {\rm O}^\dagger (2m + h_O) {\rm O}
\nn\\
&&
+ \left[ ({\rm S}^\dagger h_{SO} {\rm O} + {\rm H.C.}) + {\rm C.C.}  \right]
+ \left[ {\rm O}^\dagger h_{OO} {\rm O} + {\rm C.C.}  \right]
+ \left[ {\rm O}^\dagger h_{OO}^A {\rm O}  h_{OO}^B + {\rm C.C.}  \right]
\Biggr\}\Bigg|_{\hbox{${\bf E}=\bfPi$}},
\nn\\
&& \hspace{-8mm} H = \int d^3X \, h.
\label{HpNRQCD}
\eea
The generators ${\bf P}$ and ${\bf J}$ are exactly known:
\bea
&& \hspace{-8mm}
{\bf P} = \int d^3X \int d^3x \, {\rm Tr} \,\Biggl\{
{\rm S}^\dagger {\bf P}_X {\rm S} + {\rm O}^\dagger {\bf P}_X {\rm O} \Biggr\}
+ {1\over 2} \int d^3X \, [{\bfPi}^{a}\times, {\bf B}^{a}],
\label{PpNRQCD}
\\
&& \hspace{-8mm}
{\bf J} = \int d^3X \int d^3x \, {\rm Tr} \,\Biggl\{
{\rm S}^\dagger \left({\bf X} \times {\bf P}_X + {\bf x} \times {\bf p}_x
+ {\bfsigma^{(1)}+\bfsigma^{(2)} \over 2}  \right){\rm S}
\nn\\
&&
+ {\rm O}^\dagger \left({\bf X} \times {\bf P}_X + {\bf x} \times {\bf p}_x
+ {\bfsigma^{(1)}+\bfsigma^{(2)} \over 2}  \right) {\rm O} \Biggr\}
+ {1\over 2} \int d^3X \, {\bf X} \times [{\bfPi}^{a}\times, {\bf B}^{a}].
\label{JpNRQCD}
\eea
The Lorentz-boost generators at order $x^2/m^0$, $x^0/m$ and  $(x/m) \, {\bf
  P}_X$ are given by:
\bea
&& \hspace{-8mm} {\bf K} = -t \, {\bf P} + \int d^3X \, {1\over 2}\, \{ {\bf X}, h \}
+  \int d^3X \int d^3x \, {\rm Tr} \,\Bigl\{
  \left[ {\rm S}^\dagger {\bf k}_{SS} {\rm S} + {\rm C.C.} \right]
\nn\\
&&\qquad\qquad
+ \left[ ({\rm S}^\dagger {\bf k}_{SO} {\rm O} + {\rm H.C.}) + {\rm C.C.} \right]
+ \left[ {\rm O}^\dagger {\bf k}_{OO} {\rm O} + {\rm C.C.} \right] \Bigr\},
\label{KpNRQCD}
\\
\nn
\\
&& {\bf k}_{\varphi\phi} = {\bf k}_{\varphi\phi}^{(0,2)}
+ {\bf k}_{\varphi\phi}^{(1,-1)}
+ {\bf k}_{\varphi\phi}^{(1,0)}
+ {\bf k}_{\varphi\phi}^{(1,1)}({\bf P}_X), \qquad {\rm with}
\\
&&\qquad
{\bf k}_{\varphi\phi}^{(0,2)\, i} =
-{1\over 8} k_{\varphi\phi a}^{(0,2)}(x) \, {\bf x}^i \,({\bf x}\cdot g \bfPi)
-{1\over 8} k_{\varphi\phi b}^{(0,2)}(x) \, {\bf x}^2 \, g \bfPi^i,
\\
&&\qquad
{\bf k}_{\varphi\phi}^{(1,-1)\,i} =
- {1\over 8m} \{ k_{\varphi\phi a}^{(1,-1)}(x),
\left(\bfsigma^{(1)} \times {\bf p}_x\right)^i \}
\nn\\
&&\qquad\qquad
- {1\over 8m} \{ {k_{\varphi\phi b'}^{(1,-1)}(x) \over x^2}
\epsilon_{ij\ell} \, {\bf x}^j \, {\bf x}^k \, \bfsigma^{(1)\,k},  {\bf p}_x^\ell \}
\nn\\
&&\qquad\qquad
- {1\over 8m} \{ {k_{\varphi\phi b''}^{(1,-1)}(x) \over x^2}
\epsilon_{ijk} \, {\bf x}^\ell \, {\bf x}^k \, \bfsigma^{(1)\,j},  {\bf p}_x^\ell \}
\nn\\
&&\qquad\qquad
- {1\over 8m} \{ {k_{\varphi\phi b'''}^{(1,-1)}(x) \over x^2}
\epsilon_{\ell jk} \, {\bf x}^i \, {\bf x}^j \, \bfsigma^{(1)\,k},  {\bf p}_x^\ell \},
\\
&&\qquad
{\bf k}_{\varphi\phi}^{(1,0)\,i} =
{1\over 8m}
\{ k_{\varphi\phi a'}^{(1,0)}(x) \, {\bf x}^i, {\bf P}_X\cdot {\bf p}_x \}
\nn\\
&&\qquad\qquad
+ {1\over 8m}
\{k_{\varphi\phi a''}^{(1,0)}(x)\,{\bf x}^\ell, {\bf P}_X^\ell \, {\bf p}_x^i \}
\nn\\
&&\qquad\qquad
+ {1\over 8m}
\{ k_{\varphi\phi a'''}^{(1,0)}(x) \, {\bf x}^\ell, {\bf P}_X^i\, {\bf p}_x^\ell
\}
\nn\\
&&
\qquad\qquad
+ {1\over 8m}
\{ {k_{\varphi\phi b}^{(1,0)}(x)\over x^2} \, {\bf x}^i \, {\bf x}^\ell \,
{\bf x}^k, {\bf P}_X^\ell \, {\bf p}_x^k \}
\nn\\
&&
\qquad\qquad
- {1\over 8m} k_{\varphi\phi c}^{(1,0)}(x) \, \left(\bfsigma^{(1)}\times {\bf P}_X\right)^i
\nn\\
&&
\qquad\qquad
- {1\over 8m} {k_{\varphi\phi d'}^{(1,0)}(x)\over x^2}
\epsilon_{ij\ell} \, {\bf x}^j \, {\bf x}^k \,  \bfsigma^{(1)\,k}
{\bf P}_X^\ell
\nn\\
&&
\qquad\qquad
- {1\over 8m} {k_{\varphi\phi d''}^{(1,0)}(x)\over x^2}
\epsilon_{ijk} \, {\bf x}^\ell \, {\bf x}^k \,  \bfsigma^{(1)\,j}
{\bf P}_X^\ell
\nn\\
&&
\qquad\qquad
- {1\over 8m} {k_{\varphi\phi d'''}^{(1,0)}(x)\over x^2}
\epsilon_{\ell jk} \, {\bf x}^i \, {\bf x}^j \,  \bfsigma^{(1)\,k}
{\bf P}_X^\ell ,
\\
&& \qquad
{\bf k}_{\varphi\phi}^{(1,1)\,i}({\bf P}_X) = 0,
\eea
where $k_{\varphi\phi}^{(i,j)}$ are matching coefficients specific of ${\bf K}$.
For ${\bf k}^{(1,1)}$ only the ${\bf P}_X$-dependent terms are displayed.

The terms of ${\bf K}$ proportional to ${\bf k}_{SS}^{(0,2)}$,  ${\bf k}_{SO}^{(1,-1)}$,
and  ${\bf k}_{SO}^{(1,0)}$ vanish under color trace.
Due to charge conjugation, also the term proportional to ${\bf k}_{SO}^{(0,2)}$ vanishes.
The matching coefficients $k_{OO a}^{(0,2)}$,
$k_{\varphi\varphi a}^{(1,-1)}$, $k_{\varphi\varphi a'}^{(1,0)}$ and $k_{\varphi\varphi c}^{(1,0)}$
are equal to $1$ at tree level. All other matching coefficients, which are specific
of ${\bf K}$, are zero at ${\cal O} (\als^0)$.
Also in the case of ${\bf K}$ the tree-level matching has been performed
by multipole expanding the NRQCD Lorentz-boost generators (\ref{KNRQCD})
and projecting on singlet and octet two-particles states.
We note that loop corrections can , in principle, be calculated
as they are for the matching coefficients of the pNRQCD Lagrangian.

\subsection{Poincar\'e algebra constraints in pNRQCD}
\label{seccpn}
Let us now consider the constraints induced by the Poincar\'e algebra
(\ref{A1})-(\ref{A9}) on the pNRQCD generators $H$ and ${\bf K}$.

As in the NRQCD case, the constraint $[{\bf P}^i,{\bf K}^j] = -i \delta_{ij}
H$ has been already used in writing Eq. (\ref{KpNRQCD}).
Indeed, it forces ${\bf K}$ to have the form
$\displaystyle \int d^3X \, \{ {\bf X}, h({\bf X},t) \}/2 \,+\,$
translational-invariant terms that depend on ${\bf
  X}$ only through the canonical variables.

From $[{\bf K}^i,{\bf K}^j] = -i \epsilon_{ijk}{\bf J}^k$
at ${\cal O}(x^0/m^0)$ it follows that
\bea
&& k_{SSa'}^{(1,0)} - k_{SSa''}^{(1,0)}
 = k_{OOa'}^{(1,0)} - k_{OOa''}^{(1,0)}= 1,
\label{cQCD0}
\\
&& k_{SSc}^{(1,0)}= k_{OOc}^{(1,0)} = 1,
\\
&& k_{SSd'}^{(1,0)} = k_{OOd'}^{(1,0)} = 0,
\\
&& k_{SSd''}^{(1,0)} + k_{SSd'''}^{(1,0)} =
k_{OOd''}^{(1,0)} + k_{OOd'''}^{(1,0)} = 0,
\\
\nn
\\
&&
c_{S}^{(1,0)} = c_{O}^{(1,0)} = 1.
\label{kin}
\eea
The constraint (\ref{kin}) follows also from
$[H,{\bf K}^i] = -i {\bf P}^i$ at ${\cal O}(x^0/m^0)$.

From $[{\bf K}^i,{\bf K}^j] = -i \epsilon_{ijk}{\bf J}^k$
and $[H,{\bf K}^i] = -i {\bf P}^i$ at ${\cal O}(x/m^0)$
it follows that
\bea
&& V_{SO}^{(0,1)}= V_{SO}^{(1,1)},
\label{va1}
\\
&& V_{OO}^{(0,1)}= V_{OO}^{(1,1)}.
\label{vb1}
\eea

From $[H,{\bf K}^i] = -i {\bf P}^i$ at ${\cal O}(x^0 /m)$,
we obtain:
\bea
&& V_{SOd}^{(1,0)}= V_{SOe}^{(2,0)},
\label{va2}
\\
&& V_{OOd}^{(1,0)}= V_{OOe}^{(2,0)},
\label{vb2}
\\
&&\nn
\\
&& V_{LS\, Sa} + \left( k_{SSa}^{(1,-1)} +  k_{SSb''}^{(1,-1)} \right)
{V_S^{(0)\prime}\over 2 x} = 0,
\label{cQCD1}
\\
&& V_{LS\, Oa} + \left( k_{OOa}^{(1,-1)} +  k_{OOb''}^{(1,-1)} \right)
{V_O^{(0)\prime}\over 2x} = 0,
\label{cQCD2}
\\
&&\nn
\\
&& V_{{\bf L}^2\, Sa} + \left(k_{SSa'}^{(1,0)} + k_{SSa''}^{(1,0)} + k_{SSb}^{(1,0)}\right)
{xV_S^{(0)\prime}\over 2} = 0,
\label{cQCD3}
\\
&& V_{{\bf L}^2\, Oa} + \left(k_{OOa'}^{(1,0)} + k_{OOa''}^{(1,0)} + k_{OOb}^{(1,0)}\right)
{xV_O^{(0)\prime}\over 2} = 0,
\label{cQCD4}
\\
&&\nn
\\
&& V_{{\bf p}^2\, Sa} + V_{{\bf L}^2 Sa} + {V_S^{(0)}\over 2}
-  k_{SSa'''}^{(1,0)} {x V_S^{(0)\prime}\over2} = 0,
\label{cQCD5}
\\
&& V_{{\bf p}^2\, Oa} + V_{{\bf L}^2 Oa} + {V_O^{(0)}\over 2}
- k_{OOa'''}^{(1,0)} {x V_O^{(0)\prime}\over2} = 0,
\label{cQCD6}
\\
&&\nn
\\
&& 2\, c_F V_{SOb}^{(1,0)} -c_s V_{SOa}^{(2,0)} - k_{SSa}^{(1,-1)} V_{SO}^{(0,1)} =0,
\label{cQCD7}
\\
&& 2\, c_F V_{OOb}^{(1,0)} - c_s V_{OOa}^{(2,0)}
- {k_{OOa}^{(1,-1)} \over 2} V_{OO}^{(0,1)}
- {k_{OOc}^{(1,0)} \over 2} =0,
\label{cQCD8}
\\
&& 2\, V_{O\otimes Ob}^{(1,0)} - V_{O\otimes Oa}^{(2,0)}
- {k_{OOa}^{(1,-1)}\over 2}  V_{OO}^{(0,1)}
+ { k_{OOc}^{(1,0)}\over 2} =0,
\label{cQCD9}
\\
&& k_{SSa}^{(1,-1)} - k_{OOa}^{(1,-1)} =0,
\label{cQCDk1}
\\
&&\nn
\\
&& 2\, V_{SOc}^{(1,0)} - V_{SOb'}^{(2,0)} - k_{SSb'}^{(1,-1)} V_{SO}^{(0,1)} =0,
\label{cQCD10}
\\
&& 2\, V_{OOc}^{(1,0)} - V_{OOb'}^{(2,0)} - {k_{OOb'}^{(1,-1)} V_{OO}^{(0,1)}
  \over 2} - {k_{OOd'}^{(1,0)} \over 2} =0,
\label{cQCD11}
\\
&& 2\, V_{O\otimes Oc}^{(1,0)} - V_{O\otimes Ob'}^{(2,0)}
- {k_{OOb'}^{(1,-1)} V_{OO}^{(0,1)} \over 2} + {k_{OOd'}^{(1,0)} \over 2} =0,
\label{cQCD12}
\\
&& k_{SSb'}^{(1,-1)} - k_{OOb'}^{(1,-1)} =0,
\label{cQCDk2}
\\
&&\nn
\\
&& V_{SOb''}^{(2,0)} + k_{SSa}^{(1,-1)} \, x\, V_{SO}^{(0,1)\prime}
+ k_{SSb''}^{(1,-1)} \left(x \, V_{SO}^{(0,1)}\right)^\prime =0,
\label{cQCD13}
\\
&& V_{OOb''}^{(2,0)}
+ {k_{OOa}^{(1,-1)} \, x\, V_{OO}^{(0,1)\prime} \over 2}
+ {k_{OOb''}^{(1,-1)} \left(x \, V_{OO}^{(0,1)}\right)^\prime \over 2}
+ {k_{OOd''}^{(1,0)} \over 2}=0,
\label{cQCD14}
\\
&&
V_{O\otimes Ob''}^{(2,0)}
+ {k_{OOa}^{(1,-1)} \, x\, V_{OO}^{(0,1)\prime} \over 2}
+ {k_{OOb''}^{(1,-1)} \left(x \, V_{OO}^{(0,1)}\right)^\prime \over 2}
- {k_{OOd''}^{(1,0)} \over 2}=0,
\label{cQCD15}
\\
&& k_{SSb''}^{(1,-1)} - k_{OOb''}^{(1,-1)} =0,
\label{cQCDk3}
\\
&&\nn
\\
&& V_{SOb'''}^{(2,0)} + k_{SSb'''}^{(1,-1)} V_{SO}^{(0,1)} =0,
\label{cQCD16}
\\
&& V_{OOb'''}^{(2,0)} + {k_{OOb'''}^{(1,-1)} V_{OO}^{(0,1)} \over 2}
+  {k_{OOd'''}^{(1,0)}\over 2}=0,
\label{cQCD17}
\\
&& V_{O\otimes Ob'''}^{(2,0)} + {k_{OOb'''}^{(1,-1)} V_{OO}^{(0,1)} \over 2}
-  {k_{OOd'''}^{(1,0)}\over 2}=0,
\label{cQCD18}
\\
&& k_{SSb'''}^{(1,-1)} - k_{OOb'''}^{(1,-1)} =0,
\label{cQCDk4}
\\
&&\nn
\\
&& V_{OOa}^{(1,0)} - c_{O}^{(1,-2)} k_{OOa}^{(0,2)} + 2 \, k_{OOa''}^{(1,0)}
+ V_{OOc'}^{(2,0)} =0,
\label{cQCD20}
\\
&& V_{OOa}^{(1,0)} + c_{O}^{(1,-2)} k_{OOa}^{(0,2)} - 2 \, k_{OOa'}^{(1,0)}
- V_{OOc''}^{(2,0)}=0,
\label{cQCD21}
\\
&&  c_{O}^{(1,-2)} {\left(x^2\, k_{OOb}^{(0,2)}\right)^\prime \over x}
- 2 \, k_{OOa'''}^{(1,0)} - V_{OOc'''}^{(2,0)} =0,
\label{cQCD22}
\\
&& c_{O}^{(1,-2)} \, x \, k_{OOa}^{(0,2)\prime}
- 2\, k_{OOb}^{(1,0)} - V_{OOd}^{(2,0)} =0,
\label{cQCD23}
\eea
where $f'\equiv df/dx$. More precisely, as in the case of NRQCD discussed
in Sec. \ref{secnrqcdcons}, what we obtain are the above combinations of
matching coefficients multiplying some pNRQCD operators. For simplicity, here
we have not displayed the expressions involving the operators.
However, they may turn out to be important in considering the QED case
of the above equalities. So, for instance, the expressions appearing
in Eqs. (\ref{cQCD8}) and (\ref{cQCD9}) contribute to the
same operator  in the QED case. Therefore, the two equalities reduce
to their sum, which is given by Eq. (\ref{cQCD7}).
Same considerations apply to the couples of equations
(\ref{cQCD11})-(\ref{cQCD12}), (\ref{cQCD14})-(\ref{cQCD15}) and
(\ref{cQCD17})-(\ref{cQCD18}).
Moreover, the combinations of matching
coefficients appearing in Eqs. (\ref{cQCD20})-(\ref{cQCD23}) multiply
operators that vanish in the QED case. Therefore, these constraints
have no QED analogue. For what concerns the equalities
(\ref{va1})-(\ref{cQCD6}) in the QED case, they become pairwise identical.

\subsection{Unitary transformations}
\label{secunitary}
The generators of the Poincar\'e algebra are defined up to unitary
transformations. We can use this freedom in order to change or reduce
our basis of operators. (For a similar use of unitary transformations in a slightly
different context we refer to \cite{m1}.)
In particular, we can look for a basis where some of the matching coefficients of ${\bf K}$
are fixed. Consider the unitary transformation:
\bea
{\cal U}  &=& {\rm exp} \left(
i\int d^3X \int d^3x \, {\rm Tr} \left\{
\left[ {\rm S}^\dagger \left( {{\bf P}_X\cdot
\left(\tilde{\bf k}^{(1,-1)}_{SS} + \tilde{\bf k}^{(1,0)}_{SS}\right)\over 2m}
\right){\rm S} +  {\rm C.C.} \right]
\right.\right.
\nn\\
&& \qquad\qquad\qquad \left.\left.
+
\left[ {\rm O}^\dagger \left( {{\bf P}_X\cdot
\left(\tilde{\bf k}^{(1,-1)}_{OO} + \tilde{\bf k}^{(1,0)}_{OO}\right)\over 2m}
\right){\rm O} +  {\rm C.C.} \right] \right\} \right),
\eea
where the operators $\tilde{\bf k}^{(i,j)}_{\varphi\phi}$ are equal to the
operators ${\bf k}^{(i,j)}_{\varphi\phi}$ but with arbitrary coefficients
$\tilde{k}^{(i,j)}_{\varphi\phi}$.
The transformed Lorentz-boost generators ${\cal U} \, {\bf K} \, {\cal U}^\dagger$
have the same structure as the original ones up to order $x^2/m^0$, $x^0/m$
and  $(x/m) \, {\bf P}_X$, but with the matching coefficients shifted
as follows:
\bea
&& \hspace{-6mm} k_{\varphi\phi a}^{(1,-1)} \to k_{\varphi\phi a}^{(1,-1)} + {\tilde
    k}_{\varphi\phi a}^{(1,-1)},
\\
&& \hspace{-6mm} k_{\varphi\phi b'}^{(1,-1)} \to k_{\varphi\phi b'}^{(1,-1)} + {\tilde
    k}_{\varphi\phi b'}^{(1,-1)},
\\
&& \hspace{-6mm} k_{\varphi\phi b''}^{(1,-1)} \to k_{\varphi\phi b''}^{(1,-1)} + {\tilde
    k}_{\varphi\phi b''}^{(1,-1)},
\\
&& \hspace{-6mm} k_{\varphi\phi b'''}^{(1,-1)} \to k_{\varphi\phi b'''}^{(1,-1)} + {\tilde
    k}_{\varphi\phi b'''}^{(1,-1)},
\\
&& \hspace{-6mm} k_{\varphi\phi a'}^{(1,0)} \to k_{\varphi\phi a'}^{(1,0)} + {\tilde
    k}_{\varphi\phi a'}^{(1,0)} + {\tilde k}_{\varphi\phi a''}^{(1,0)},
\\
&& \hspace{-6mm} k_{\varphi\phi a''}^{(1,0)} \to k_{\varphi\phi a''}^{(1,0)} + {\tilde
    k}_{\varphi\phi a'}^{(1,0)} + {\tilde k}_{\varphi\phi a''}^{(1,0)},
\\
&&  \hspace{-6mm}
k_{\varphi\phi a'''}^{(1,0)} \to k_{\varphi\phi a'''}^{(1,0)} + 2\, {\tilde
    k}_{\varphi\phi a'''}^{(1,0)},
\\
&& \hspace{-6mm} k_{\varphi\phi b}^{(1,0)} \to k_{\varphi\phi b}^{(1,0)} + 2\, {\tilde
    k}_{\varphi\phi b}^{(1,0)},
\\
&& \hspace{-6mm} k_{\varphi\phi d''}^{(1,0)} \to k_{\varphi\phi d''}^{(1,0)} + {\tilde
    k}_{\varphi\phi d''}^{(1,0)} - {\tilde k}_{\varphi\phi d'''}^{(1,0)},
\\
&&  \hspace{-6mm}
k_{\varphi\phi d'''}^{(1,0)} \to k_{\varphi\phi d'''}^{(1,0)} - {\tilde
    k}_{\varphi\phi d''}^{(1,0)} + {\tilde k}_{\varphi\phi d'''}^{(1,0)}.
\eea
We have, now, the freedom to choose the coefficients
$\tilde{k}^{(i,j)}_{\varphi\phi}$. A convenient choice is that one
that fixes the above matching coefficients to their tree level values:
\be
k_{\varphi\phi a}^{(1,-1)}, k_{\varphi\phi a'}^{(1,0)} \to 1,
\quad
k_{\varphi\phi b'}^{(1,-1)}, k_{\varphi\phi b''}^{(1,-1)}, k_{\varphi\phi
  b'''}^{(1,-1)}, k_{\varphi\phi a''}^{(1,0)}, k_{\varphi\phi a'''}^{(1,0)},
k_{\varphi\phi b}^{(1,0)}, k_{\varphi\phi d''}^{(1,0)}, k_{\varphi\phi d'''}^{(1,0)}
\to 0.
\label{basis}
\ee
The unitary transformation leaves invariant the generators ${\bf P}$ and ${\bf
  J}$, while the changes in ${\cal U}\, H \, {\cal U}^\dagger$
may be reabsorbed in a redefinition of the matching coefficients.

Finally, we note that other unitary transformations are possible
and some of them may, in principle, further reduce the number
of operators in ${\bf K}$ or $H$. We will not examine this issue
here, which, however, deserves further studies.

\subsection{Discussion}
\label{secexact}
We comment now on the obtained relations.
Equation (\ref{kin}) fixes the centre-of-mass kinetic energy to be
equal to ${\bf P}_X^2/4m$. It may come as a surprise that, at least
at the order at which we are working here,  Poincar\'e
invariance does not fix the coefficient of the kinetic energy ${\bf
  p}_x^2/m$ of the quarks in the centre-of-mass frame.
However, one should consider that the coordinate ${\bf x}$ is no more a
dynamical variable of the theory. Nevertheless, we may argue
that, because no other momentum-dependent operator
than the kinetic energy of NRQCD, $- \psi^\dagger \, \bfnabla^2/(2m) \, \psi +
\chi^\dagger \, \bfnabla^2/(2m) \, \chi$, may contribute to the kinetic energy of
pNRQCD, the coefficients $c_{S}^{(1,0)}$ and $c_{S}^{(1,-2)}$
have to be equal. It follows then that also
$c_{S}^{(1,-2)}=1$ (analogously for $c_{O}^{(1,-2)}$).

From Eqs. (\ref{cQCD1}) and (\ref{cQCD2}) we obtain:
\be
{V_{LS\, Sa} \over V_S^{(0)\prime}} = {V_{LS\, Oa} \over V_O^{(0)\prime}}
= -\frac{1}{2x},
\label{ex0}
\ee
where the last equality holds in the basis of operators discussed in Sec. \ref{secunitary}.
Eq. (\ref{ex0}) in the singlet sector
is the relation between the spin-orbit potentials
and the static potential first derived in \cite{poG}
by boosting the potentials expressed in terms of Wilson loops.
It was also obtained in \cite{bgv}.
In all these previous cases the Lorentz-boost generators were written
in the basis of Eq. (\ref{basis}).
The extension to the octet sector is new.

From Eqs.  (\ref{cQCD3})-(\ref{cQCD6}) we obtain:
\bea
&& V_{{\bf L}^2\, Sa} + {x \, V_S^{(0)\prime} \over 2}
=
V_{{\bf L}^2\, Oa} + {x \, V_O^{(0)\prime} \over 2}
= 0,
\label{ex0bis}
\\
&& V_{{\bf p}^2\, Sa} + V_{{\bf L}^2\, Sa} + {V_S^{(0)} \over 2}
=
V_{{\bf p}^2\, Oa} + V_{{\bf L}^2\, Sa} + {V_O^{(0)} \over 2}
= 0.
\label{ex0tris}
\eea
These equations hold in the basis of Sec. \ref{secunitary}.
In the singlet sector they are the relations between the
momentum-dependent potentials first derived in \cite{poBBP} by boosting the potentials
expressed in terms of Wilson loops. They were also  obtained in \cite{bgv}.
The extension to the octet sector is new.

Equations (\ref{va1})-(\ref{vb2}) constrain the fields to
enter in the singlet-octet and octet-octet sectors of the Lagrangian
just in the combination
\be
{\bf x}\cdot \left(g{\bf E} +
{1\over 2} \left\{ {{\bf P}_X\over 2 m} \times, g {\bf B}\right\} \right),
\ee
i.e. like in the Lorentz force.
The coefficients  $V_{SO}^{(0,1)}$ and $V_{OO}^{(0,1)}$ were called
$V_A$ and $V_B$ respectively in the previous literature \cite{static,pnrqcd}.
These are the coefficients associated to the next-to-leading order terms
of the multipole expansion of the static pNRQCD Lagrangian.
They play an important role in the running of the static potentials
\cite{static,RGstatic} and in several observables \cite{decay}.
Eqs. (\ref{cQCD7})-(\ref{cQCDk4}) mix them with the matching coefficients
of operators appearing at order $1/m$ and $1/m^2$.
Eqs. (\ref{cQCD13})-(\ref{cQCD15}) involve also the derivatives of
$V_{SO}^{(0,1)}$ and $V_{OO}^{(0,1)}$.
Eqs. (\ref{cQCD7}) and (\ref{cQCD8}) contain combinations of
matching coefficients inherited from NRQCD. Somehow these
relations reflect at the level of the pNRQCD potentials the relation
(\ref{constrNRQCD3}) among the NRQCD matching coefficients.

The four sets of Eqs. (\ref{cQCD7})-(\ref{cQCDk1}), (\ref{cQCD10})-(\ref{cQCDk2}),
(\ref{cQCD13})-(\ref{cQCDk3}) and (\ref{cQCD16})-(\ref{cQCDk4}) can be
combined in order to give the following four equations that involve only
potentials appearing in the Lagrangian:
\bea
&& \hspace{-6mm} {2\, c_F V_{SOb}^{(1,0)} -c_s V_{SOa}^{(2,0)} \over  V_{SO}^{(0,1)}}
=  {2\, \left(c_F V_{OOb}^{(1,0)} + V_{O\otimes Ob}^{(1,0)}\right)
- \left( c_s  V_{OOa}^{(2,0)} + V_{O\otimes Oa}^{(2,0)}\right) \over
V_{OO}^{(0,1)}}
= 1,
\label{ex1}
\\
&& \hspace{-6mm} {2\, V_{SOc}^{(1,0)} - V_{SOb'}^{(2,0)} \over  V_{SO}^{(0,1)}}
= {2\, \left(V_{OOc}^{(1,0)} + V_{O\otimes Oc}^{(1,0)} \right)  -
\left(V_{OOb'}^{(2,0)} + V_{O\otimes Ob'}^{(2,0)} \right) \over
V_{OO}^{(0,1)}}
= 0,
\label{ex2}
\\
&& \hspace{-6mm} {-V_{SOb''}^{(2,0)} + 2\, c_F V_{SOb}^{(1,0)} - c_s V_{SOa}^{(2,0)} \over
\left(x \, V_{SO}^{(0,1)}\right)^\prime }
=
\nn\\
&& \hspace{2mm}
{- V_{OOb''}^{(2,0)} - V_{O\otimes Ob''}^{(2,0)}
+ 2\, \left(c_F V_{OOb}^{(1,0)} + V_{O\otimes Ob}^{(1,0)}\right)
- \left( c_s  V_{OOa}^{(2,0)} + V_{O\otimes Oa}^{(2,0)}\right) \over
\left(x \, V_{OO}^{(0,1)}\right)^\prime}
= 1,
\label{ex3}
\\
&& \hspace{-6mm} { V_{SOb'''}^{(2,0)} \over V_{SO}^{(0,1)}}
= {V_{OOb'''}^{(2,0)} + V_{O\otimes Ob'''}^{(2,0)} \over V_{OO}^{(0,1)}}
= 0,
\label{ex4}
\eea
where the last equalities hold in the basis of Sec. (\ref{secunitary}).

Eqs.  (\ref{cQCD20})-(\ref{cQCD23}) are typical for the non-Abelian structure of QCD
and have been commented above. Here, we add that, summing Eq. (\ref{cQCD20}) and
(\ref{cQCD21}), we obtain:
\be
V_{OOa}^{(1,0)} =1 + {V_{OOc''}^{(2,0)} - V_{OOc'}^{(2,0)} \over 2}.
\label{ex5}
\ee

\section{Conclusions}
\label{seccon}
In this work we have shown how the implementation of the Poincar\'e
algebra provides a feasible and useful tool to constrain the
dynamics of non-relativistic effective field theories of QCD.
The method was used in the past, at a quantum-mechanical level,
to constrain the relativistic dynamics of some phenomenological potentials.
This is its first use in an effective field theory context.

In Sec. \ref{secnrqcd} we have applied the method to the NRQCD/HQET
Lagrangian up to order $1/m$ and $(\bfnabla^2 \,\bfnabla^i) /m^2$
in the commutators. In a simple way we have obtained some of
the relations, Eqs. (\ref{constrNRQCD1})-(\ref{constrNRQCD4}),
derived formerly from reparametrization invariance.
This shows that reparametrization invariance is, indeed,
one way in which the Poincar\'e invariance of QCD manifests itself in
the HQET (for a similar observation in a different context see \cite{eiras}).
An obvious extension of this work would be
to study the constraints induced by Poincar\'e invariance
at higher order in $1/m$. This study would be of interest, because
at some point the four-fermion operators of NRQCD will start to play a role
and new relations, specific of NRQCD, will show up.
To our knowledge, relativistic invariance of NRQCD,
in the sector where it does not reduce to the HQET, has never been explored.
We believe that the framework used here is suitable also for this exploration.

In Sec. \ref{secpnrqcd} we have calculated the constraints induced
in pNRQCD by the Poincar\'e invariance of QCD  up to order $x/m^0$
and $x^0 /m$. We have obtained a set of new constraints,
listed in Eqs. (\ref{cQCD0})-(\ref{cQCD23}). These constraints
involve, in general, mixings of matching coefficients appearing
in the pNRQCD Lagrangian and in the Lorentz-boost generators.
There are two kinds of information that we can extract from them.
First we can combine them in order to obtain relations that
involve only the potentials appearing in the pNRQCD Lagrangian.
These equations are (\ref{kin})-(\ref{vb2}),
(\ref{ex0})-(\ref{ex0tris}) and (\ref{ex1})-(\ref{ex5}).
More relations of this kind are expected in going to higher orders.
In some specific bases of operators the relations become particularly
simple. In the present letter we have presented a unitary
transformation that fixes the matching coefficients specific of ${\bf K}$
at their tree-level value. An exhaustive analysis was beyond the purposes
of the present letter but surely deserves further investigations.
A second information, provided by the constraints, is the form of the
Lorentz-boost generators expressed in terms of the potentials appearing
in the Lagrangian of pNRQCD. Once ${\bf K}$ is known, the singlet and octet
quark-antiquark fields transform under infinitesimal Lorentz boosts
with velocity ${\bf v}$ in accordance to $\delta S = i [S,{\bf v}\cdot{\bf K}]$ and
$\delta O^a = i [O^a,{\bf v}\cdot{\bf K}]$.

Finally, we would like to mention that the present approach
is quite general. Therefore, it may be also suited to derive exact
relations among the matching coefficients
of other effective field theories of QCD, where
the manifest covariance under boosts has been destroyed by an expansion
in some small momenta. An example is the soft-collinear
effective theory \cite{scet}, where also reparametrization invariance
has been discussed \cite{repscet}.

\vspace{0.3cm}

\noindent
{\bf Acknowledgements.}
The authors thank Heinz Rothe and Klaus Rothe for valuable discussions on the
quantization procedure and Joan Soto for reading the manuscript
and for useful comments and suggestions.
A.V.  was supported during this work by the European Community through the
Marie-Curie fellowship HPMF-CT-2000-00733. N.B. gratefully acknowledge
the support of the Alexander von Humboldt Foundation.

\vfill\eject

\end{document}